\documentclass[prl,aps,superscriptaddress,twocolumn]{revtex4-1}

\usepackage{tabularx}
\usepackage{amsmath}
\usepackage{amssymb}
\usepackage{graphicx}
\usepackage{dcolumn}
\usepackage{bm}
\usepackage{ulem}
\usepackage{color}

\begin{document}

\title{Towards scalable entangled photon sources with self-assembled InAs/GaAs quantum dots}
\author{Jianping Wang}
\affiliation{Key Laboratory of Quantum Information, University of Science and Technology of China, Hefei, 230026, China}
\affiliation{Synergetic Innovation Center of Quantum Information and Quantum Physics, University of Science and Technology of China, Hefei, 230026, China}
\author{Ming Gong}
\affiliation{Department of Physics and Center for Quantum Coherence, The Chinese University of Hong Kong, Shatin, N.T., Hong Kong, China}
\author{G-C Guo}
\affiliation{Key Laboratory of Quantum Information, University of Science and Technology of China, Hefei, 230026, China}
\affiliation{Synergetic Innovation Center of Quantum Information and Quantum Physics, University of Science and Technology of China, Hefei, 230026, China}
\author{Lixin He}
\email{Email address: helx@ustc.edu.cn}
\affiliation{Key Laboratory of Quantum Information, University of Science and Technology of China, Hefei, 230026, China}
\affiliation{Synergetic Innovation Center of Quantum Information and Quantum Physics, University of Science and Technology of China, Hefei, 230026, China}

\begin{abstract}
{\bf Biexciton cascade process in self-assembled quantum dots (QDs) provides an
  ideal system for deterministic entangled photon pair source
  \cite{Benson00,Shields07}, which
 is essential in quantum information science.
  The entangled photon pairs have recently be realized in
  experiments~\cite{Stevenson06,bennett10,Trotta14}
  after eliminating the FSS of exciton using a number of
  different methods. However, so far the QDs entangled photon sources are not scalable,
  because the wavelengths of the QDs are different from dot to dot.
  Here we propose a wavelength tunable entangled photon
  emitter on a three dimensional stressor, in which the FSS and exciton energy
  can be tuned independently, allowing photon entanglement between
  dissimilar QDs.  We confirm these results by using
  atomistic pseudopotential calculations. This provides a first step towards
  future realization of scalable entangled photon generators for quantum
  information applications.
}
\end{abstract}

\maketitle

Entangled photon pairs
play a crucial role in quantum information
applications, including
quantum teleportation\cite{jennewein02}, quantum
cryptography\cite{gisin02} and
distributed quantum computation\cite{cirac99}, etc.
The biexciton cascade process in a self-assembled QD has been proposed
\cite{Benson00} to generate the ``event-ready'' entangled photon pairs.
As shown in Fig.~1(a), a biexciton decays into two photons via two paths of different
polarizations $|H\rangle$ and $|V \rangle$. If the two
paths are indistinguishable, the final result is a
polarization entangled photon pair state\cite{Stevenson06,
Benson00}$(|H_{XX} H_{X}\rangle +  |V_{XX} V_{X}\rangle) /\sqrt{2}$.
However, the $|H\rangle$-
and $|V\rangle$-polarized
photons have a small energy difference,
known as the fine structure splitting (FSS), which
is typically about -40 $\sim$ +80 $\mu$eV in the InAs/GaAs QDs
\cite{young05,hogele04,tartakovskii04}, much larger than the
radiative linewidth ($\sim$ 1.0 $\mu$eV)
\cite{Stevenson06, Akopian06}.
Such a splitting provides therefore ``which way'' information about the photon
decay path that can destroy the photon entanglement,
leaving only classically correlated photon pairs
\cite{Stevenson06,Akopian06}.
Great efforts have been made trying to eliminate the FSS of excitons in QDs,
and significant progress has been made in understanding \cite{bester03,he08,singh10,gong11}
and manipulating the FSS in
self-assembled QDs in recent years. Various techniques has been developed to
eliminate the FSS in
QDs\cite{seidl06,Gerardot07,vogel07,dou08,ding10,bennett10,jons11,Trotta12}.
Especially, it was recently found by applying
combined uniaxial stresses or stress together with electric field, it is possible to
reduce to the FSS to nearly zero for general self-assembled InAs/GaAs QDs
\cite{Wang12, Trotta12, Trotta14}.

However, to build practical QDs
devices for applications in quantum information science, they must be scalable.
%
One possible application for scalable entangled photon emitters
is shown in Fig. 1(b) as quantum repeater to
distribute entanglement over long distance. The set-up of Fig. 1(b) can also
be used to generate multi-photon entanglement \cite{Pan12,Huang11a}.
The on-demand entangled photon
emitters have great advantages of
over the traditional parametric down convention process to generate
multi-photon entanglement, which has finite probability of generating
more than one photon pair in a excitation cycle\cite{gisin02}.
In these applications, the
wavelengths of the joint photons have to be identical,
i.e., $\lambda_2$=$\lambda_3$ in Fig. 1(b).
Besides, one often need to interface the entangled photon pairs to other
quantum system, such as NV-center, cold atom, or other solid quantum systems
etc.
These applications also requires that the wavelengths of the QDs to be tunable,
while at the same time keep the FSS nearly zero.
However, it was found that there are strong correlations
between exciton energy and FSS of exciton\cite{Gerardot07, Pooley14}.
Furthermore, because of the random alloy distribution
and other uncontrollable effects,
the physical properties of QDs differ dramatically from dot to dot.
Therefore, it is still a great challenge to build such
scalable entangled photon generators using dissimilar quantum dots.

\begin{figure}
\centering
\includegraphics[width=3.2in]{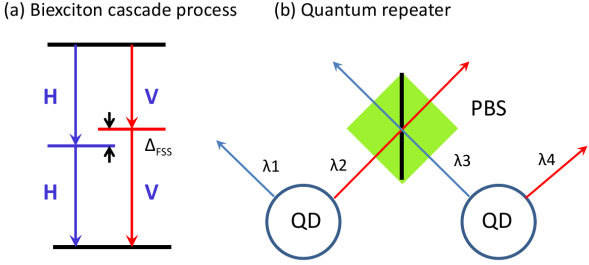}
\caption{{\bf Scalable entangled photon pairs from QDs}.
{\bf (a)} A schematically show of
biexciton cascade process. The energy difference between the H and V polarized photons
is known as the
fine structure splitting (FSS). To have entangled photon pairs, FSS must be smaller than 1 $\mu$eV.
{\bf (b)} QDs entangled photon emitters
used in a quantum repeater. The entangled photon pairs from the two QDs
are entangled by the PBS, which requires $\lambda_2$=$\lambda_3$.
This set-up can also be used to generate multi-photon entanglement.}
\label{fig-fig1}
\end{figure}

The independent tunability the FSS and exciton energy is therefore
essential for the scalable entangled photon emitters. We demonstrate such
a tunability by proposing a three dimensional stressor for QDs.
Our basic setup is schematically shown Fig. 2 (a).
We consider QDs that are tightly glued to the $yz$ plane of
the piezoelectric lead zirconic titanate (PZT) ceramic
stack~\cite{seidl06}.
The [100] axis of the QDs samples are aligned to the polar
($z$) axis of PZT, whereas
[010], [001] axes of the QDs are aligned to the $y$ and $x$ axes
of the PZT respectively.
Two independent in-plane electric voltages,
$V_z$ and $V_y$, are applied to the PZT device as shown in the Fig. 2(a), which
generate electric fields $F_z$ and $F_y$ along the PZT $z$ and $y$ axes
respectively.
The electric field causes the in-plane strain to the QDs as,
\begin{eqnarray}
\tensor{e} =
\begin{pmatrix}
d_{33}  & 0  &  0  \\
0  & d_{31}  &  0  \\
  0  &  0     &  d_{\perp}
\end{pmatrix} F_z
+
\begin{pmatrix}
0  & d_{15}  &  0 \\
d_{15}  & 0  &  0 \\
  0  & 0       &  0
\end{pmatrix} F_y,
\end{eqnarray}
where, $d_{33}$, $d_{31}$, $d_{15}$ are the piezoelectric coefficients of
PZT and $d_{\perp}=(d_{33}+d_{31})S_{12}/(S_{11}+S_{12})$.
$S_{11}$, $S_{12}$ and $S_{44}$ are the elastic
compliance constants of GaAs.
The electric fields $F_z$ and $F_y$ lead to in-plane strains to the QDs
are shown in Fig. 2(b): $F_z$ causes strain along [010] and [100] axes of the
QDs sample,
whereas $F_y$ causes strain along [110] axis of the QDs.
As demonstrated in Ref.~\cite{Wang12}, one
can almost fully eliminate the FSS in a general InAs/GaAs QD by
suitable combination of such strains.
To tune the energy of the exciton, we apply a stress along the [001]
direction of the QDs sample [see Fig. 2(b)], which can be easily implemented
in experiments. This pressure generates the strain $e_{zz}$ to the QDs.
Now we have a device that can tune freely the 3D strain to the
QDs. Next we show that the device is able to tune the exciton emission energy
in a wide range while keep the FSS minimum ($<$ 0.1 $\mu$eV).

\begin{figure}
\centering
\includegraphics[width=3.2in]{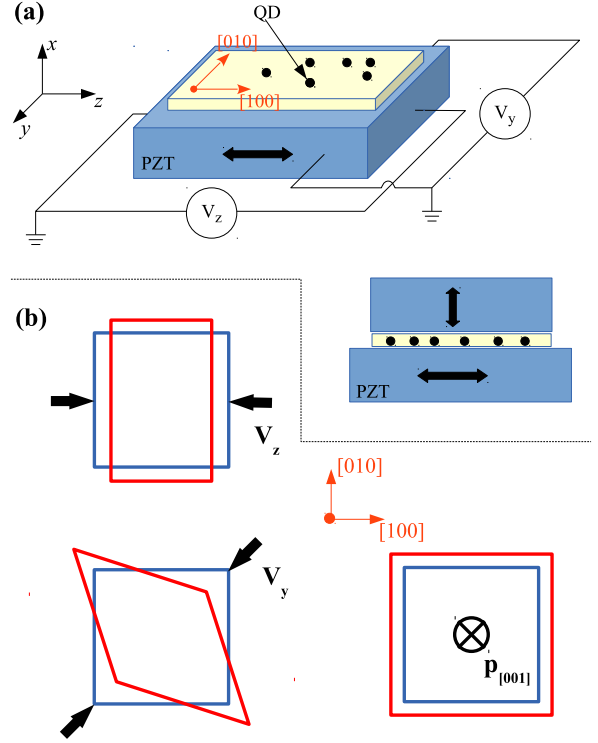}
\caption{{\bf Basic setup for wavelength tunable entangled photon pairs}. {\bf (a)}
  A three dimensional stressor that can tune the FSS and exciton energy
  independently in QDs. {\bf (b)} The two bias voltages $V_z$ and $V_y$ are applied
  to generate in-plane strain, which is used to tune the FSS of exciton.
  The $p_{[001]}$ is used to tune the exciton energy.
  The blue and red structures represent the shapes of QDs
  before and after applied voltages and stress respectively.}
\label{fig-fig2}
\end{figure}

\begin{figure}
\centering
\includegraphics[width=3.2in]{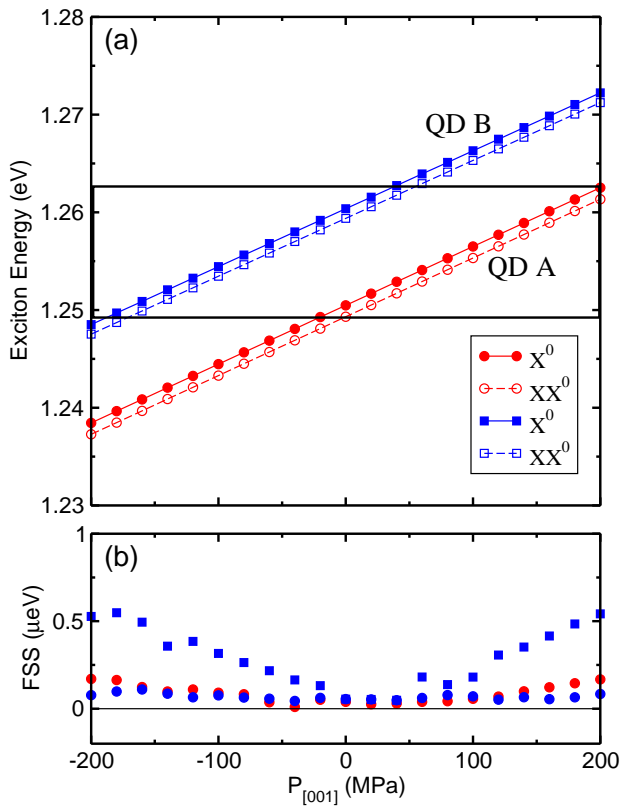}
\caption{{\bf Tuning of exciton and biexciton energies and FSS.}
  {\bf (a)} The exciton and biexciton energies of QD A and QD B as functions of
  stress $p_{[100]}$, under fixed $F_z$ and $F_y$. {\bf (b)} The FSS
  of QD A and QD B as functions of stress $p_{[100]}$.
  The blue squares are the FSS of QD B
  with $F_z$= 3.5 kV/cm and $F_{y}$=4.3 kV/cm, whereas the blue dots
  are the FSS of QD B after further optimizing $F_{y}$.
}
\label{fig-fig3}
\end{figure}

To see if our device really works,
we perform atomistic pseudopotential calculations (see Methods) to confirm
the above predictions. We have calculated 8 (In,Ga)As/GaAs dots. The details of the
structure and alloy composition are given in Table S4 of the Supplementary
materials~\cite{supplement}.
The results of two dots QD-A and QD-B are presented in Fig. 3 (a).
The results are obtained in such way: First, in the absence of
$p_{[001]}$, we carefully choose the in-plane electric fields $F_z$ and $F_y$
to tune strain tensor $\tensor{e}$,
that reduces the FSS of exciton to nearly zero~\cite{Wang12}.
For QD-A, the applied in-plane
electric fields are $F_{z}({\rm  A})$=9.6 kV/cm,
and $F_{y}({\rm A})$=3.3 kV/cm, whereas for QD-B, the electric fields
are $F_{z}({\rm  B})$=3.5 kV/cm,
and $F_{y}({\rm B})$=4.3 kV/cm, respectively.
We then switch on the perpendicular stress to study the evolution of
exciton energy and FSS as functions of $p_{[001]}$.
Figure 3(a) depicts the exciton and biexciton emission energies
for QD-A and QD-B as functions $p_{[001]}$, while keep the in-plane
electric field $F_{z}$ and $F_{y}$ (thus the in-plane strain) unchanged.
Although, in practice, one can only apply positive (compression)
pressure to the QDs in our device,
we plot the results of negative pressure just for theoretical interest.
We find that the exciton energy can be tuned in a wide
range of about 20 meV when $p_{[001]}$ change from -200 MPa to 200 MPa, with the
slope of $\sim$ 6 meV/100 MPa for both QDs.
The change of exciton energy is comparable with the full width at the half maximum of a general QDs
ensemble. These results suggest that in principle, the exciton
energies of most QDs grown in the same sample can be tuned to identical using our scheme.
The corresponding results for FSS are presented in Fig. 3(b).
Remarkably, the FSS change with $p_{[001]}$ is rather small.
For QD-A, the FSS [the red dots in Fig. 3(b)] is about 0.03 $\mu$eV at $p_{[001]}$=0. It become slightly
larger with the increasing of $p_{[001]}$, and reach $\sim$ 0.1 $\mu$eV at
$p_{[001]}$=$\pm$ 200 MPa.
The FSS of QD-B [the blue squares in Fig. 3(b)]  has somehow
stronger dependence of $p_{[001]}$, which reaches
approximately 0.5 $\mu$eV at $p_{[001]}$=$\pm$ 200 MPa. This is nevertheless
still smaller than the homogeneous broadening of the spectral ($\sim$ 1
$\mu$eV), which is the upper limit for entangled photon generation.
In this situation, it is possible to further reduce the FSS at given
$p_{[001]}$, by tuning the in-plane electric fields $F_z$ and $F_y$. The blue dots
are the FSS of QD-B  after such optimization.
By slightly changing
$F_{y}({\rm  B})$ from 4.3 kV/cm to 4.5 kV/cm,
the FSS reduces from approximately 0.5 $\mu$eV to approximately 0.08 $\mu$eV
at $p_{[001]}$=200 MPa.
This change will shift the exciton energy by only about 0.02 meV.
This energy shift can be compensated by increasing
$p_{[001]}$ by 0.36 MPa, which hardly change the FSS. In such way, we can tune the
FSS to nearly zero at any given exciton energy in the range in only one
or two iterations.
We also calculate the exciton radiative lifetimes under $p_{[001]}$. The exciton
lifetimes for QD A and QD B are around 1 ns, and change little under
$p_{[001]}$, which is good for the proposed device
applications.

More results for dots with different geometries and alloy compositions
are given in Table S5 of the supplementary materials~\cite{supplement}.
We fit the atomic pseudopotential calculated results by a 2$\times$2 model
\cite{gong11,Wang12}.
Although it is easy to understand that in-principle the FSS and exciton energy can
be tuned simultaneously to desired values by suitable combination of three linearly independent external fields
from the 2$\times$2 model,
there is an additional advantage that in our scheme
the exciton energy and FSS can be
tuned almost separately, i.e., the in-plane strain
have very strong effects on the FSS, and
relatively small effect to the exciton energy. In contrast, $p_{[001]}$ have
strong effect on the exciton energy, but rather small effect to the FSS.
The (nearly) independent tuning of FSS and exciton energy is an enormous
advantage for the scalable entangled photon sources. The electric field may
also be used to tune the FSS \cite{bennett10,Trotta12}. However,
at the same time the exciton energies change
dramatically under electric field due to the stark effects.
It is therefore harder to tune
both quantities to the target values,
which requires to tune the three external fields simultaneously.

Now we try to understand the above results in several different levels.
First we would like to understand
why in-plane strains have  small effects on
$E_X$, but $p_{[001]}$ have large effect on $E_X$?
Because the envelope functions of the electron and hole
states change little if the external strain is not very large,
the direct electron-hole Coulomb interaction also change
little (See Figure S1 in the supplementary materials\cite{supplement}).
The change of exciton energy is therefore
mainly determined by the single-particle energies gap $E_g$. We can
estimate the slope of exciton emission energy (or recombination energy)
to the stress as,
\begin{equation}
\frac{dE(X^0)}{dp}\approx\frac{dE_g}{dp}\, .
\end{equation}
If we neglect the $O(p^2)$ terms, the slope of band gap under
the stress along the [001] direction can be written as according to the
Bir-Pikus model\cite{supplement},
\begin{equation}
  \frac{d E_{g}}{dp}
 \approx -a_g (S_{11}+2 S_{12})- b_{v}
  (S_{11}-S_{12}) \,.
\end{equation}
For the in-plane stresses
along the [010], [100] and [110] directions, we have,
\begin{equation}
  \frac{d E_{g}}{dp}
  \approx -a_g (S_{11}+2 S_{12})+\frac{1}{2} b_{v}
  (S_{11}-S_{12}).
  \label{eq:p110}
\end{equation}
Here $a_g$=$a_c$-$a_v$=-6.08 eV is the deformation potential for band gap,
and $a_c$, $a_v$ are the deformation potentials for the conduction band,
and valence bands respectively. $b_v$=-1.8 eV is the biaxial deformation potential
of the valence bands.
Because of the cancelation between the first term and the second
term in Eq.~\ref{eq:p110}, the in-plane stresses have small effects on the band gap.
On the other hand, the stress along the [001] direction has
much larger impact on the exciton energy because the first term adds up to
the second term.

The second question is why the in-plane stresses (strain)
have more important influence on FSS than the [001] stress (strain)?
Intuitively, as shown Fig.~\ref{fig-fig2}(b), $F_z$ and $F_y$ change the in-plane
anisotropy of the QDs, whereas $p_{[001]}$ does not.
The microscopic mechanism of strain tuning of FSS in self-assembled InAs/GaAs QDs has been
studied in Ref.~\onlinecite{Wang14}, where some of us derived analytically the change of
FSS of excitons under the external stresses using the Bir-Pikus model.
For simplicity, we illustrate the results using a 6$\times$6 model. We have,
\begin{equation}
\Delta_{\rm FSS}=2|K_{\mathrm{od}}| \approx |2(\kappa+i\delta) + 4\varepsilon_+ K| \, ,
\end{equation}
where $K_{\mathrm{od}}$ is the off-diagonal element of exchange integral
matrix, equivalent to half the FSS. $\kappa$, $\delta$ and $K$ are exchange
integrals over different orbital functions\cite{Wang14}. Especially, 2$K \sim$
300 -- 400 $\mu$eV is approximately the dark-bright exciton energy splitting.
The exchange integrals over different orbital
functions only changes slightly under external strain. The change of FSS is
mainly due to the bands mixing~\cite{Wang14},
\begin{equation}\label{eq:epsilon}
      \varepsilon_+
      =\frac{R^{\ast}}{2\sqrt{3}}\left(\frac{1}{Q}+\frac{9}{\Delta}\right)
        +\frac{3(S^\ast)^2}{2Q\Delta} \, ,
\end{equation}
where $R$, $Q$, $\Delta$, $S$ are parameters in  Bir-Pikus model (See Supplementary
materials\cite{supplement}).
As one can see from Eq.~\ref{eq:epsilon}, $Q$ only appears in the denominator and has a much larger
value than $R$ and $S$, therefore
the change of $\varepsilon_+$ under stress mainly depends on the
slope of $R$ and $S$.
As shown in Table S6 of the Supplementary materials\cite{supplement},
the stress along the [001] direction only changes isotropic and biaxial
strains, i.e. only change $Q$, therefore have little effect on the slope of
$\varepsilon_+$. On the other hand, the in-plane stresses modify the
in-plane anisotropy of the QDs, i.e., $e_{xx}$-$e_{yy}$,
which changes the $R$, and therefore modifies HH-LH coupling and the FSS
\cite{supplement}.

To conclude, we proposed a novel portable device that
allow to tune the FSS and exciton
energies of (In,Ga)/GaAs QDs (nearly) independently.
This provides a first step
towards future realization of scalable entangled photon pairs generators
for quantum information applications, such as long distance entanglement
distribution, multi-phonon entanglement and interfaces to other quantum
systems, ect.
The device can be implemented using current experimental techniques.

The authors thank C-F Li, J-S Xu, Y-F Huang and B-S Shi
for valuable discussions.
LH acknowledges the support from the Chinese National
Fundamental Research Program 2011CB921200, the
National Natural Science Funds for Distinguished Young Scholars, and
the support from the Central Universities funding WK2470000006.
M.G. is supported by Hong Kong RGC/GRF Projects (No. 401011 and No. 2130352), University Research
Grant (No. 4053072) and The Chinese University of Hong Kong (CUHK) Focused Investments Scheme.

\vskip 0.5cm
{\bf METHODS}

We model the InAs/GaAs quantum dots by embedding the InAs dots into a
60$\times$60$\times$60
8-atom GaAs supercell. The QDs are assumed to be grown along the [001]
direction,
on the top of the one monolayer InAs wetting layers\cite{gong08a}.
To calculate the exciton energies and their FSS, we first have to obtain
the single-particle energy levels and wavefunctions
by solving the Schr\"{o}dinger equation,
\begin{equation}
  \left[ -\frac{1}{2} \nabla^2
+ V_{\rm ps}({\bf r}) \right] \psi_i({\bf r})
=\epsilon_i \;\psi_i({\bf r}) \; ,
\label{eq:schrodinger}
\end{equation}
where $V_{\rm ps}({\bf r}) = V_{\rm SO}+ \sum_n\sum_{\alpha} v_{\alpha}({\bf r}
- {\bf R}_{n,\alpha})$ is the superposition of
local screened atomic pseudopotential{\bf s} $v_{\alpha}({\bf r})$,
and the total (non-local) spin-orbit (SO) potential $V_{SO}$.
The atom positions $\{ {\bf R}_{n,\alpha} \}$ of type $\alpha$ at site $n$
are obtained by minimizing the total strain energies due to the dot-matrix
lattice mismatch using the valence
force field (VFF) method \cite{keating66}.
The pseudopotentials of the InAs/GaAs QDs are taken
from Ref. \onlinecite{williamson00}, which have been well
tested.
The Schr\"{o}dinger equations are
solved via a Linear Combination of Bulk Bands (LCBB)
method~\cite{wang99b}.

The exciton energies are calculated via the many-particle
configuration interaction (CI) method \cite{franceschetti99},
in which the (many-particle) exciton wavefunctions are
expanded in Slater determinants for single and biexcitons constructed
from all of the confined single-particle electron
and hole states.
The exciton energy is obtained by diagonalizing the full Hamiltonian in
the above basis, where
the Coulomb and exchange integrals are computed numerically
from the pseudopotential single-particle states, using
the microscopic position-dependent dielectric constant.
Including spin, this state is fourfold degenerate.
The electron-hole Coulomb interactions leave this fourfold degeneracy intact.
The FSS arises from the asymmetric electron-hole exchange
matrix\cite{bester03}.
The piezo-effects were ignored in the calculation, as it was shown in
Ref. \onlinecite{ediger07b} that the FSS does not change much in the InAs/GaAs
QDs by including the piezo-effects.


%

\end{document}